\documentclass[10pt, letterpaper]{IEEEtran}

\usepackage{color}
\usepackage{url}
\usepackage{hyperref}
\usepackage[latin9]{inputenc}
\usepackage{epstopdf}
\usepackage{amsthm}
\usepackage{amsmath}
\usepackage{amsthm}
\usepackage{amssymb}
\usepackage{graphicx}
\usepackage{subfigure}
\usepackage{wrapfig}
\usepackage{algorithmic}
\usepackage{algorithm}
\usepackage{mathrsfs}
\usepackage{float}
\usepackage{mathtools}
\usepackage{tabulary}
\usepackage{booktabs}
\usepackage{caption}
\usepackage{multirow}
\usepackage{array}
\usepackage{makecell}
\usepackage{graphicx}

\usepackage{caption}
\usepackage[export]{adjustbox}
\usepackage{tabularray}
\PassOptionsToPackage{bookmarks=false}{hyperref}

\IEEEoverridecommandlockouts
\usepackage[dvipsnames]{xcolor}

\usepackage{graphicx}
\usepackage{caption}
\usepackage[export]{adjustbox}
\usepackage{tabularray}

\usepackage[letterpaper,
            left=0.7in,
            right=0.7in,
            top=0.75in,
            bottom=1.2in]{geometry}

\def\BibTeX{{\rm B\kern-.05em{\sc i\kern-.025em b}\kern-.08em
   T\kern-.1667em\lower.7ex\hbox{E}\kern-.125emX}}

\newcommand{\comment}[1]{ }
\newcommand{\TPD}{{\tt{TPD}}}

\newcommand\subparagraph{%
  \@startsection{subparagraph}{0}
  {\parindent}
  {0ex \@plus 0ex \@minus 0ex}
  {-1em}
  {\normalfont\normalsize\bfseries}}
\makeatother

\usepackage{titlesec}
\titlespacing*{\section}      {0pt}{*1.5}{*0.7} 
\titlespacing*{\subsection}   {0pt}{*1.2}{*0.6} 
\titlespacing*{\subsubsection}{0pt}{*1.0}{*0.3} 

\begin{document}


 


\title{Bluetooth Fingerprint Identification Under Domain Shift Through Transient Phase Derivative}

\author{Haytham Albousayri, Bechir Hamdaoui, Weng-Keen Wong, Nora Basha\\
\small School of Electrical Engineering and Computer Science, Oregon State University, Corvallis, OR, USA\\
Emails: \{albousah, hamdaoui, wongwe, bashano\}@oregonstate.edu 
}

\IEEEaftertitletext{\vspace{-2em}}
\maketitle

\thispagestyle{empty}
\pagestyle{empty}

\begin{abstract}
Deep learning-based radio frequency fingerprinting (RFFP) has become an enabling physical-layer security technology, allowing device identification and authentication through received RF signals. This technology, however, faces significant challenges when it comes to adapting to domain variations, such as time, location, environment, receiver and channel. For Bluetooth Low Energy (BLE) devices, addressing these challenges is particularly crucial due to the BLE protocol's frequency-hopping nature.
In this work, and for the first time, we investigated the frequency hopping effect on RFFP of BLE devices, and proposed a novel, low-cost, domain-adaptive feature extraction method. Our approach improves the classification accuracy by up to 58\% across environments and up to 80\% across receivers compared to existing benchmarks.

\end{abstract}



\section{Introduction}

Despite its widespread adoption, Bluetooth Low Energy (BLE) remains susceptible to a range of security threats~\cite{barua2022security}. Adversaries can exploit vulnerabilities in device pairing, data transmission, and connection protocols, risking the exposure of sensitive personal data and the disruption of critical applications~\cite{sevier2019analyzing, zhang2020breaking}. Recently, deep learning-based radio frequency (RF) fingerprinting (RFFP) has gained attention as a promising approach to mitigate these risks~\cite{zhang2019physical, shen2023deep}. RFFP leverages hardware-induced imperfections in RF signals to extract unique, device-specific fingerprints, enabling automatic and robust device identification.

\begin{figure}
    \centering
    \includegraphics[width=\linewidth]{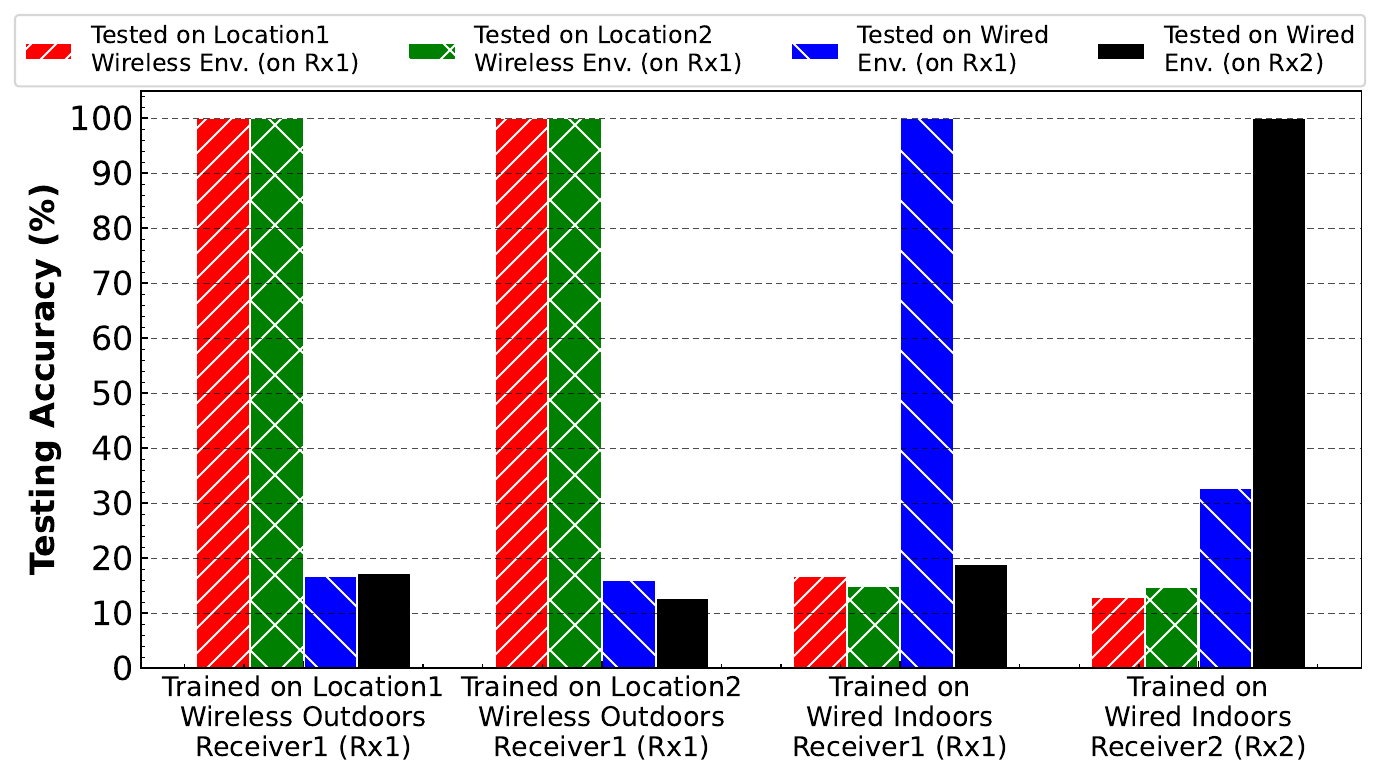}
    \caption{Testing classification accuracy for 31 BLE devices across different environments and different receivers. More on this experiment/testbed can be found in Sec.~\ref{sec:Testbed}}
    \label{fig:motivation}
\end{figure}

Previous studies~\cite{elmaghbub2021lora, fu2023deep} have shown that deep learning models using raw I/Q data often suffer from poor generalization across different domains and environments, underscoring the need for more robust approaches. As illustrated in Fig.~\ref{fig:motivation}, device classification accuracy drops significantly when training and testing are conducted in mismatched conditions (e.g., different channels or receivers), indicating that models tend to capture environment-specific rather than device-specific features. To address this, we propose a novel method designed to be resilient to domain variations. Our results demonstrate strong generalization, with classification accuracy improvements of up to $58\%$ across diverse environments and up to $80\%$ across different receivers.

\subsection{Related Work}

Early research in RF fingerprinting (RFFP) primarily focused on demonstrating the feasibility of using signal characteristics to distinguish between network devices, particularly those using LoRa and WiFi technologies~\cite{shen2023toward, elmaghbubdistinguishable2024, jagannath2022rf}. For example, Elmaghbub et al.~\cite{elmaghbubdistinguishable2024} investigated the impact of domain shifts on WiFi and LoRa fingerprinting and proposed a method that leverages the signal power spectrum envelope to mitigate domain variability. 
Jagannath et al.~\cite{jagannath2022rf} addressed domain shift using short-time Fourier transforms combined with an attention-based classifier. Additionally, Shen et al.~\cite{shen2023toward} focused on the LoRa protocol and proposed data augmentation strategies to improve performance under low SNR conditions.


In the context of Bluetooth, RF fingerprinting remains underexplored, and existing studies often focus on narrow use cases. Establishing consistent fingerprints for Bluetooth devices is particularly challenging due to the protocol's frequency-hopping behavior. For example, Givehchian et al.~\cite{givehchian2022evaluating} explored a location-tracking attack by estimating hardware impairments in BLE devices. Ali et al.~\cite{ali2019assessment} were among the first to examine BLE fingerprinting, using the Hilbert transform on the transient portion of Bluetooth Basic Rate signals from 20 mobile phones to enhance classification across varying SNRs. Jagannath et al.~\cite{jagannath2023embedding} tackled domain adaptation in 10 BLE devices, but their method struggled to generalize across domains due to limited data and a lack of frequency-hopping mitigation. More recently, Kashani et al.~\cite{kashani2024radio} studied BLE-based RF fingerprinting in Wireless Body Area Networks, employing a complex-valued CNN to classify 13 on-body sensors.

While existing techniques have achieved modest gains in testing accuracy, they often fall short of offering practical or scalable solutions for real-world deployment. Several key limitations in current research remain unaddressed. A primary challenge stems from the frequency-hopping behavior of BLE, which introduces a domain adaptation problem: if a classifier is trained on device fingerprints captured at a specific frequency (e.g., Channel 1), can it still accurately identify those devices when they transmit on different channels? Moreover, the generalizability of these techniques across varied environments remains uncertain. Most studies also restrict their evaluations to fewer than 20 devices, whereas real-world applications demand scalability to support significantly larger device populations. Bridging these gaps is crucial for advancing BLE RF fingerprinting and enabling its adoption in practical, large-scale scenarios.

\subsection{Key Contributions}
This work introduces a domain-adaptive technique that remains robust against changes in frequency channels, indoor/outdoor settings, and packet contents. Our key contributions are as follows:

\begin{itemize}
    \item \textit{\textbf{Novel data representation:}} 
We propose a novel data representation technique that effectively addresses domain variability. Despite its simplicity, our approach consistently outperforms several recently introduced methods.

    
    \item \textit{\textbf{Comprehensive BLE dataset:}} We collected a comprehensive set of BLE frames from 31 IoT devices under different environments, different receivers and different frequency channels. Both the dataset and the code implementation are publicly available at:~\href{https://research.engr.oregonstate.edu/hamdaoui/sites/research.engr.oregonstate.edu.hamdaoui/files/release_note_datasets_ble_august2025_v1.pdf}{BLE RF Fingerprint Dataset}.
    
    \item \textbf{\textit{Adaptation to frequency hopping:}} To the best of our knowledge, we are the first to experimentally evaluate the impact of frequency hopping on BLE RFFP. 
    Our results demonstrate that our method is robust to domain shifts induced by frequency hopping.
    
    \item \textbf{\textit{Adaptation to environment changes:}} Our findings demonstrate that the proposed method delivers robust RFFP performance under environmental shifts, maintaining high classification accuracy even when training is conducted indoors and testing occurs outdoors, or when different receivers are used.

\end{itemize}

\noindent The remainder of this paper is organized as follows: Section~\ref{sec:BLE} examines hardware limitations and signal variations inherent to BLE devices that contribute to device fingerprinting. Section~\ref{sec:Proposed} presents the proposed method. Section~\ref{subsec:hw-impact} illustrates the effectiveness of the proposed frame work in capturing key RF impairments. Section~\ref{sec:Testbed} describes the experimental setup and data collection method. Section~\ref{sec:results} presents the results. Finally, Section~\ref{sec:conclusion} summarizes the paper's main findings.

\section{BLE Hardware Impairments}
\label{sec:BLE}

\subsection{Ideal GFSK Modulation}
Given the baseband line-coded NRZ data $d(t)$ and the impulse response of a Gaussian pulse shaping filter defined as~\cite{linz1996efficient} $h(t) = \frac{\sqrt{\pi}}{a} e^{- {\pi ^2 t^2}/{a^2}}$ where
$a = \frac{1}{BT} \sqrt{{ln(2)}/{2}}$ is a parameter related to $3$-dB bandwidth, $B$ represents the $3$-dB bandwidth, and $T$ is the symbol duration. The Gaussian pulse-shaped stream can be expressed as $g(t) = d(t) * h(t)$, where $*$ represents the convolution operation.
The baseband angle-modulated signal $x(t)$ is\cite{linz1996efficient}
$x(t) = x_I(t) + jx_Q(t) = \cos(\phi(t)) + j \sin(\phi(t))$
with $x_I(t)$ and $x_Q(t)$ representing the In-phase (I) and the Quadrature (Q) components of the transmitted baseband signal, respectively, and $\phi (t) = 2 \pi f_m \int_{0}^{t} {g}(t) dt$ being the instantaneous angular shift function where $f_{m}$ is the peak frequency deviation.
After up-conversion with a carrier frequency $f_c$,  the ideal passband signal becomes
$$ S_{RF}(t) = A \cos(2\pi f_c t) \cos(\phi(t)) - \!A\sin(2\pi f_c t)\sin(\phi(t))$$
On the receiver side, IQ components are obtained by first multiplying the received signal with two orthogonal sinusoidals with frequency $f_c$ and then passing it through a low-pass filter. After normalization, it yields an in-phase baseband signal $y_I(t) = \cos(\phi(t))$ and a quadrature baseband signal $y_Q(t) = \sin(\phi(t))$.

\comment{
At the receiver side, IQ components are obtained by first multiplying the received signal with two orthogonal sinusoidals with frequency $f_c$ and then passing it through a low pass filter, thus yielding (after normalization):
\small
\begin{align*}
    y_I(t) &= LP(S_{RF}(t) \times \cos(2\pi f_c t))= x_I(t) = \cos(\phi(t)) \normalsize \\
    \small
    y_Q(t) &= LP(S_{RF}(t) \times -\sin(2\pi f_c t)) =x_Q(t) = \sin(\phi(t))
\end{align*} \normalsize
Where $y_I(t)$, $y_Q(t)$ and $LP(\cdot)$ are the In-phase baseband signal at the receiver side, the Quadrature baseband signal at the receiver side, and the Low Pass filter operator, respectively. 
And finally the ideal received baseband signal $y(t)$ can be be represented as $y(t) = y_I(t) + jy_Q(t)$.
}

\subsection{Impaired GFSK Modulation}
When considering hardware imperfections, each stage in the RF chain on both the transmitter and the receiver can contribute to the overall distortion of the received signal. For instance, the imperfect bandwidth duration product, denoted by $\widetilde{BT}$, can be off its optimum value due to some shifts in the 3-dB frequency in the Gaussian filter. This results in a distorted pulse-shaped waveform, $\tilde{g}(t)$. Another source of distortion can arise from the distorted version of the peak frequency deviation, modeled as $\tilde{f}_m = f_m + \Delta f$ with $\Delta f$ being the peak frequency deviation offset from its nominal value $f_m$.
%
Finally, the overall received distorted baseband signal after normalization can be expressed as~\cite{albousayri2025neural} \\$\tilde{y}(t) = \left[\tilde{y}_I(t) + j\tilde{y}_Q(t)\right]e^{j\left(2\pi f_\text{CFO}  t + \theta_{PO}\right)}$ with
%
\begin{align*}
\tilde{y}_I(t) = (1-IQ_{Amp})\cos(\phi(t) - {IQ_{Phase}}/{2}) + I_{DC}\\
\tilde{y}_Q(t) = (1+IQ_{Amp})\sin(\phi(t) + {IQ_{Phase}}/{2}) +Q_{DC}
\end{align*}
where $f_\text{CFO}$, $\theta_{PO}$, $I_{DC}$, $Q_{DC}$, $IQ_{Phase}$ and $IQ_{Amp}$ represent the carrier frequency offset, phase offset, In-phase component of DC-offset, Quadrature component of DC-offset, phase imbalance and amplitude imbalance between the In-phase and Quadrature, respectively. Detailed definitions and impacts of these impairments are discussed in Section~\ref{subsec:hw-impact}.

\section{\TPD{}: the Feature Extraction Approach}
\label{sec:Proposed}

Our experiments show that key impairments, such as CFO and phase noise, are captured more accurately when extracted from the transient and preamble portions of the received signals. Estimating these impairments from the transient and preamble portions yields more robust results, and accordingly, our technique focuses on extracting impairment-specific features from such portions.

We propose to calculate and use the derivative of the phase of the transient and preamble portions (i.e., the Transient and Preamble Phase Derivative or \TPD{} representation) as the BLE device features. To find the \TPD{} representation for a captured frame, we first sample the signal up to time $L$, the time needed to transmit the transient and preamble portion. 
We then estimate the instantaneous phase $\sigma (t)$ as $\sigma (t) = \text{unwrap} (\angle{\tilde{y}(t)})$ for any $t \in [0, L]$.
%
Finally, \TPD{} is obtained by taking the time derivative of $\sigma (t)$; i.e., $\text{\TPD{}}{(t)} \triangleq \frac{d}{dt} \sigma (t)$, which, for negligible  IQ imbalances, can be estimated as
\begin{equation}
    \text{\TPD{}}(t) \approx  2\pi {f}_{\text{CFO}} +  \frac{d \theta_{PO}}{dt}
    + 2\pi \tilde{f}_m \tilde{g}(t)
\end{equation}
Our rationale is that the derivative masks the channel effect by eliminating the phase offset; in other words, the second term will be zero in the case of a static phase offset, leaving only the device-specific impairments.\\
\begin{figure}
\centerline{\includegraphics[width=1\linewidth]{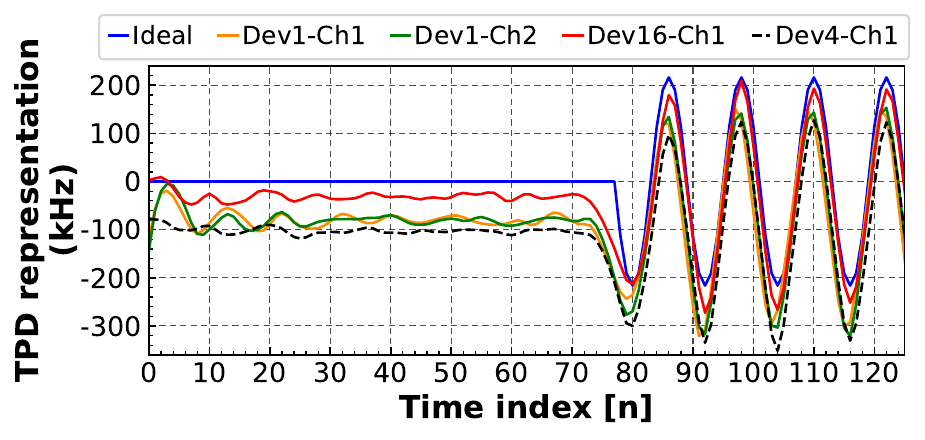}}
\caption{\TPD{} representation extracted from signals sent by different devices across different channels in the same wired setup}
\label{fig:Diff_Devices}
\end{figure}
Figure~\ref{fig:Diff_Devices} shows how different devices can have different \TPD{} representations. The figure also shows that signals from the same device but captured at different channels have similar representations, indicating that the \TPD{} representation is channel-agnostic technique.

\subsection{Discrete Time Domain \TPD{}}
\label{sub: TPD discrete}
\begin{figure}
    \centering
    \includegraphics[width=1\linewidth]{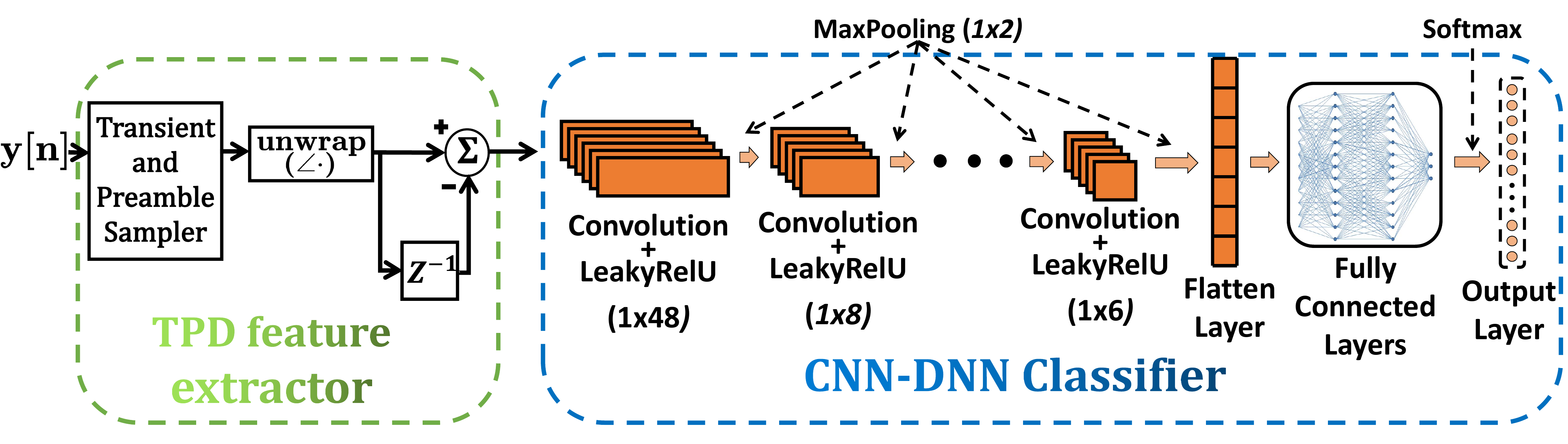}
    \caption{The block diagram of the proposed technique.}
    \label{fig:end to end}
\end{figure}

Denote by $\mathbf{y}[n] \in \mathbb{C}^N$ the discrete-time raw IQ frame, where $N = f_S \; T_{\text{Frame}}$ is the total number of discrete-time samples with $f_S$ and $T_{\text{Frame}}$ being the sampling frequency and the physical layer frame duration, respectively. Knowing that the transient and preamble parts are contained in the first $L$ samples ($L<N$), we only use the first $L$ samples of the signal for training.
We then compute the vector $\boldsymbol{\sigma}[n] \in  \mathbb{R}^L$ as $\boldsymbol{\sigma}[n] =\text{unwrap}(\angle \; \mathbf{y}[n])$.
Next, a discrete time difference is obtained to get the final representation using
$\text{\TPD{}}[n] = \boldsymbol{\sigma}[n] - \boldsymbol{\sigma}[n-1]$.
Figure~\ref{fig:end to end} presents the block diagram of the proposed technique. 

\subsection{Motivation Behind Using \TPD{} Instead of Raw IQ}
\label{sub: TPD motivation}
Beyond highlighting device-specific features and suppressing channel information (Section~\ref{subsec:hw-impact}), this approach also helps prevent overfitting to signal content. Training a deep learning RF model on the entire raw signal can easily overfit to the signal's content, making it too difficult for a classifier trained on raw IQ data with a specific protocol data unit (PDU) to accurately identify the device through other transmitted PDUs. The challenge is even greater for BLE devices, as the PDU header includes frequency-hopping information that changes dynamically during ongoing connections. 
Driven by these facts, it is necessary to find a common ground in which every single transmission from all of the devices has the same baseband content but is still affected by the hardware impairments. For that, we focus on the transient part of the signal along with the preamble of the BLE frame. 
In general, using the preamble information for RF fingerprinting is sufficient for the following reasons: 
 
\noindent {{\bf 1)}} For any given sampling rate, the dimension of each training data input must be fixed, which is crucial when dealing with time series. This cannot be ensured when using the entire frame content, such as when training using raw IQ data, since the payload size can vary.
    
\noindent {{\bf 2)}} The classifier will not be confused because of a change in the frame's content. For example, if a classifier is trained to identify $K$ devices using some PDUs, it may fail to recognize a device later using other transmitted PDUs. Our experiments suggest that this failure occurs because the classifier's overfitting to the transmitted content. In theory, to train the Raw IQ-based classifier on all possible payload contents---considering that BLE's PDU (in data channels) can range from 2 to 257 bytes of data, including the header and MIC~\cite{bluetooth60}---the minimum number of signals to be captured per device to cover all the possible cases is $\sum_{P=2}^{257}2^{P*8} \approx 8.31 \times 10^{618}$.
    
    
\noindent {{\bf 3)}} It could be computationally expensive to train using entire raw IQ frames. For instance, a 257-byte PDU frame transmitted over a 1M PHY setting and sampled at the Nyquist rate of 2 MS/s (twice the BLE channel bandwidth after down conversion) results in $4240 \times2$ (time) samples.
\begin{figure*}
    \centering
    \subfigure[CFO]
    {\includegraphics[width=0.191\linewidth]{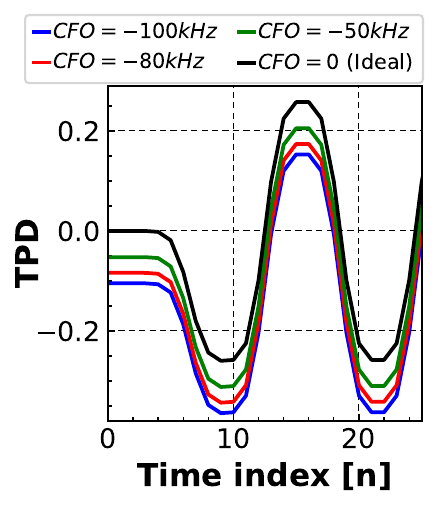}\label{sfig:CFO}}
    \subfigure[IQ Imbalance]{\includegraphics[width=0.192\linewidth]{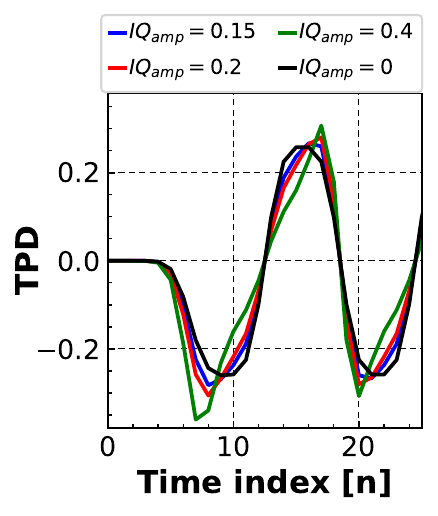}\label{sfig:IQs}}
    \hspace{-10pt}
    \subfigure[IQ Offset]
    {\includegraphics[width=0.215\linewidth]{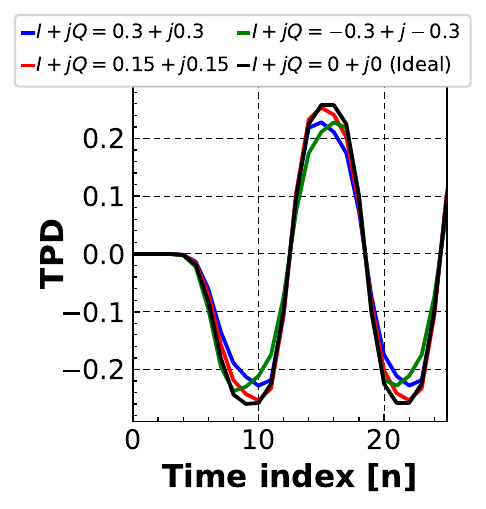}\label{sfig:DC}}
    \hspace{-10pt}
    \subfigure[MFD]{\includegraphics[width=0.192\linewidth]{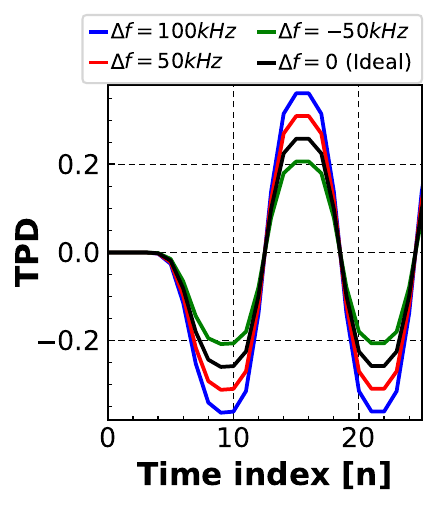}\label{sfig:MFD}}
    \subfigure[BT]{\includegraphics[width=0.198\linewidth]{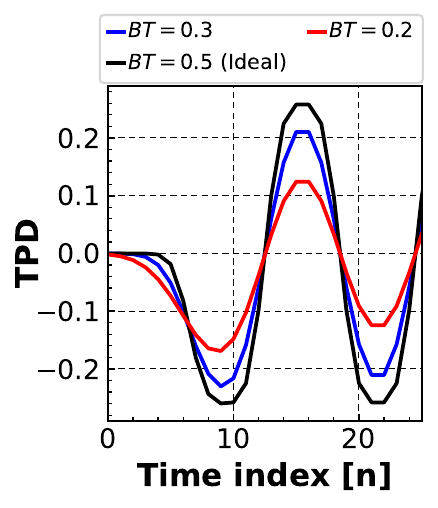}\label{sfig:BTs}}
    \caption{\TPD{} representation under the effect of hardware impairments}
    \label{fig:HWI}
\end{figure*}
\section{\TPD{}'s Effectiveness in Capturing BLE Hardware Impairments}\label{subsec:hw-impact}
To assess the robustness of \TPD{} in capturing BLE hardware impairments, we take a closer look at the \TPD{} representation of BLE signals under the effect of different impairments in isolation from each other.
For this, we use Python to simulate the proposed method, outlined in Section~\ref{sec:Proposed}, to generate BLE baseband signals capturing the impairment model described in Section~\ref{sec:BLE}. Simulations were performed at a sampling rate of 6 MS/s, following the GFSK 1M PHY specifications provided in~\cite{bluetooth60}. 
An arbitrary bit sequence was chosen for the payload, while the preamble was set to ${01010101}$. The transient part was defined as 6 samples in length, equivalent to the duration of one bit. This configuration resulted in $L=54$ samples, containing both the transient and preamble parts. However, for clarity and to avoid redundancy in visualization, only the initial $26$ samples are presented. In this demonstration, we studied the following impairments.

\subsection{Carrier Frequency Offset $(f_{CFO})$}
The Carrier Frequency Offset (CFO), a crucial parameter in modulation analysis, represents the deviation of the carrier frequency from its nominal value, caused by factors like local oscillator inaccuracy, clock rate error, Doppler shifts, and other hardware imperfections. 
Figure~\ref{sfig:CFO} presents the impact of different CFO values on the \TPD{} representation of an IQ signal. Observe that a non-zero CFO value introduces a vertical shift in the \TPD{} representation, resulting in non-overlapping data points and a clear separation between RF signals, thus illustrating the impact of different CFO values. 
In contrast, when examining the effect of CFO on raw IQ data samples, the separation is much less distinct, due to the inherent periodicity of sinusoidal signals.  



\subsection{IQ Imbalance ($IQ_{Amp}$ and $IQ_{Phase}$)}
IQ imbalance arises when there is an amplitude mismatch ($IQ_{Amp}$) and/or phase mismatch ($IQ_{Phase}$) between the in-phase (I) and quadrature (Q) components. This imperfection can result from hardware limitations, component mismatches, and/or manufacturing variations~\cite{schuchert2001novel}.
Figure~\ref{sfig:IQs} shows how the amplitude imbalance affects the \TPD{} representation, leading to sharper transitions in the pulses. For negative imbalance values, the slope of these transitions is reversed. 
Like in the case of CFO, this demonstration shows that the proposed \TPD{} captures the impact of the IQ imbalance and therefore can extract device features that are specific to the IQ imbalance.


\subsection{IQ Offset ($I_{DC}$ and $Q_{DC}$)}
In modulation schemes, maintaining a precise alignment of the I and Q components is essential for optimal signal reception. IQ offset, occurring when the IQ origin shifts from its intended position, often appears as a constant DC offset or carrier feedthrough in the modulated signal. This misalignment is typically caused by hardware imperfections, such as imbalances in the analog and digital processing chains, and can degrade signal quality if not properly corrected. Figure~\ref{sfig:DC} illustrates the effect of varying DC Offset values on our proposed \TPD{} representation. The figure shows that different DC Offset values introduce distinct distortions, highlighting the representation's capability in capturing DC Offset impairments. 
\subsection{Maximum (Peak) Frequency Deviation ($f_m$) }
Maximum Frequency Deviation (MFD) in a frequency modulator represents the difference between the maximum positive frequency and the central frequency, determined by the modulation index. In BLE, the modulation index is defined to be between 0.45 and 0.55, with an ideal value of 0.5, resulting in a frequency deviation of 250 kHz~\cite{bluetooth60}. 
However, distortion can arise due to hardware imperfections. 
We assess the impact of $\Delta f = \tilde{f}_m - f_m$, the difference between the ideal and distorted peak frequency deviation, on \TPD{} by measuring and plotting in Figure~\ref{sfig:MFD}, the \TPD{} representation under different $\Delta f$ values.
%
The figure demonstrates that different $\Delta f$ values result in different \TPD{} representations and that larger $\Delta f$ errors result in a larger swing in the \TPD{} representation. This again clearly shows the \TPD{}'s ability in capturing device signatures that are caused by MFD.



\subsection{Bandwidth Duration (BT) Product}
The BT parameter controls the smoothness of the Gaussian filter and is ideally set to 0.5~\cite{bluetooth60}. However, in real systems,
the filter's 3-dB bandwidth (B) may deviate from this ideal value, resulting in a minor deviation from the ideal BT, which is shown to be a strong device identifier~\cite{zhang2025radio}. 
In Figure~\ref{sfig:BTs}, we illustrate the impact of varying BT values on \TPD{}. Observe that smaller BT values result in lower preamble peaks and a slower transition from the transient state to the preamble.  

\subsection{Phase Offset ($\theta_{PO}$)}
\begin{figure}[h]
    \centering
    \includegraphics[width=\linewidth]{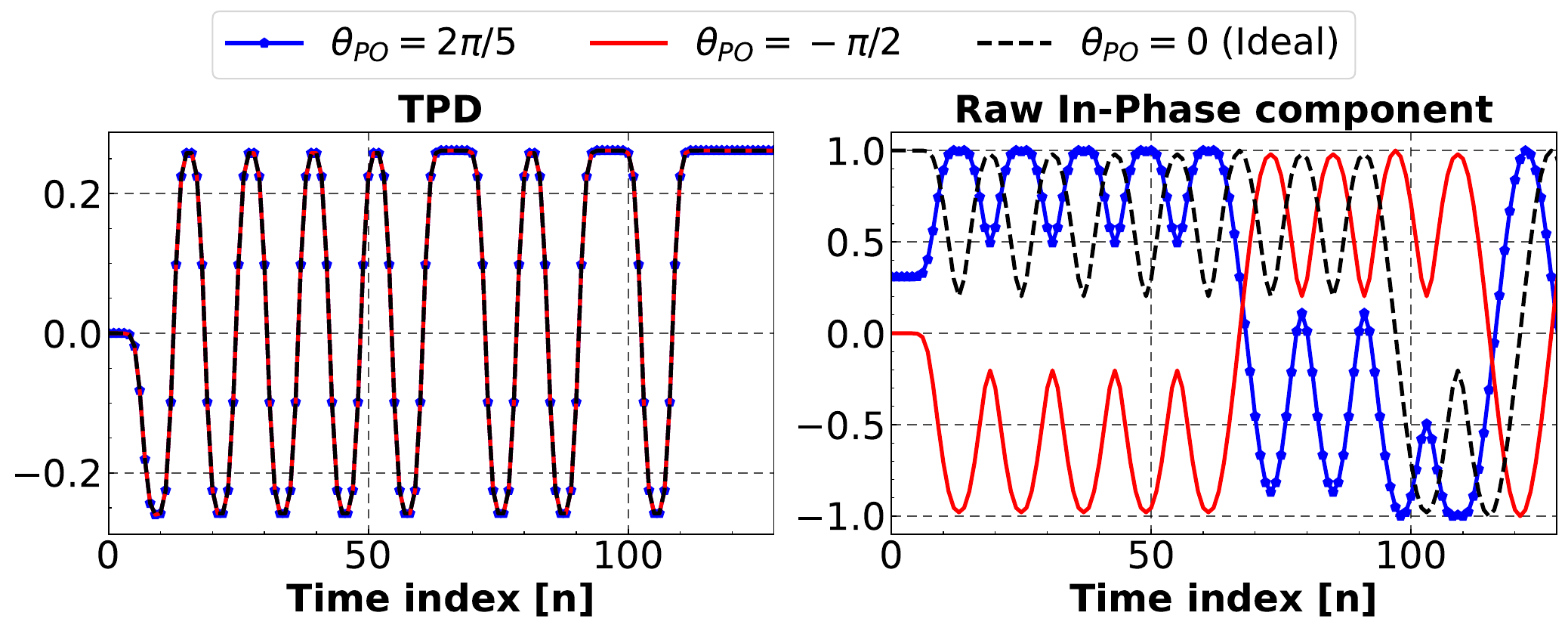}
    \caption{The impact of $\theta_{PO}$ on \TPD{} and Raw IQ}
    \label{fig:POs}
\end{figure}

Although hardware impairments due to local oscillator imperfections can cause a phase offset in the transmitted signal, the communication channel itself significantly influences the received signal phase. In time-invariant, line-of-sight (LoS) narrow-band channels, this effect can be represented as~\cite{tse_viswanath_2005} $h = \alpha e^{j \theta_{PO}}$, where $\alpha$ denotes the path loss and $\theta_{PO}$ is the phase offset. The impact of $\alpha$ can be effectively mitigated through power normalization. The phase offset $\theta_{PO}$ between the transmitted and received signals is given by $\theta_{PO} = \theta_{0} - 2\pi f_{c}d/c$, where $\theta_{0}$ represents the phase error resulting from oscillator imperfections, $f_c$ is the carrier central frequency, $d$ is the distance between the transmitter (Tx) and receiver (Rx), and $c$ is the speed of light. Consequently, the phase offset is highly sensitive to variations in both the BLE frequency channel ($f_c$) and the environmental conditions, due to its direct dependence on frequency and Tx-Rx separation. 
Figure~\ref{fig:POs} illustrates that although Raw IQ data can be significantly affected by varying values of $\theta_{PO}$ resulting from diverse channel conditions, \TPD{} effectively maps these signals into a unified representation, providing a robust, channel-agnostic framework.

\section{Testbed and Dataset Collection}
\label{sec:Testbed}



To validate our proposed technique, we conducted three experiments, namely: \textit{A)} Different environments on the same channel, \textit{B)} Different channels under the same environment, and \textit{C)} Different receivers on the same channel/environment. Throughout, the term "channel" is used to refer to BLE frequency channel.
For all of these experiments, we collected multiple datasets from 31 different Seeed Studio XIAO devices. Seeed Studio XIAO is an IoT mini-development board based on the Espressif ESP32-C3 WiFi/Bluetooth dual-mode chip. Two Ettus USRP (Universal Software Radio Peripheral) B210 receivers were employed to sample and collect the RF data in the form of raw IQ values via GNURadio, see figure~\ref{fig:outdoorsBLE}(b),~\ref{fig:outdoorsBLE}(c). 
Before starting the collection process, each device is powered on and allowed a 6-minute warm-up period to ensure hardware stabilization~\cite{elmaghbubdistinguishable2024}, then followed by a 2-minute period of data collection. The bandwidth was set to 2MHz, the sampling rate was set to 6MS/s and the power gain was set to 29dB and 8dB for the wireless data collection and wired data collection, respectively. We utilized 1M PHY modulation setup for this experiment, where each symbol represents 1 bit of information without employing any coding scheme.
The frequency channels we used for the above mentioned experiments are: Channel 1 (Ch1), Channel 2 (Ch2), Channel 14 (Ch14) and Channel 32 (Ch32), where each one of these channels is centered at 2.406GHz, 2.408GHz, 2.434GHz and 2.470GHz, respectively. 

\subsection{Different Environments on the Same Channel}
\label{sub:DESC}
\begin{figure}
    \centering
    \begin{minipage}{0.6\linewidth} 
        \centering
        \includegraphics[width=\linewidth]{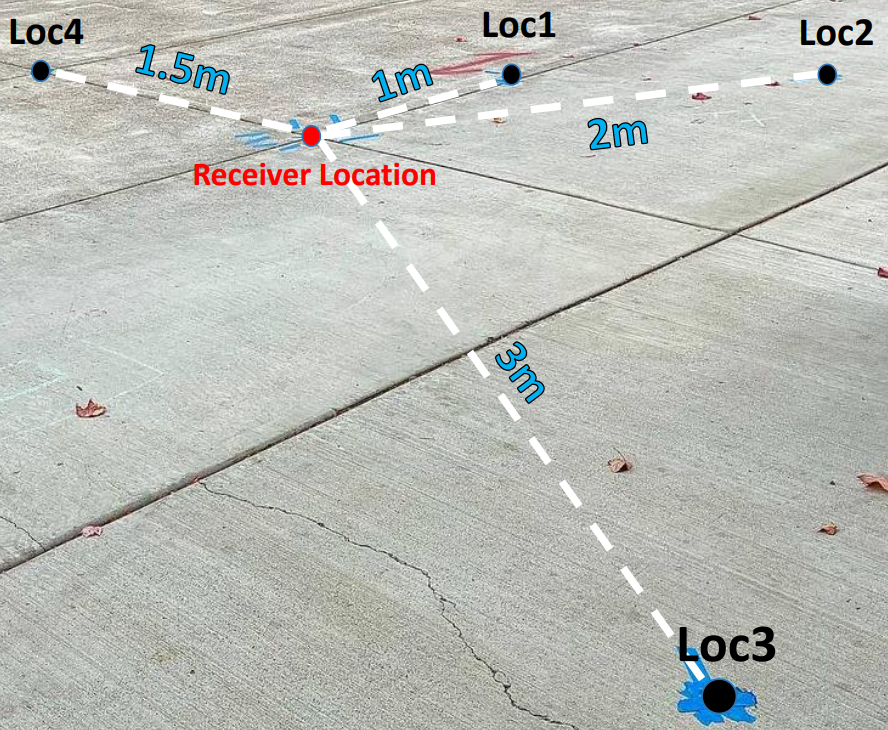}
        \SetCell[c=1]{c} {\small \text{(a) Outdoor setup}}
        \label{fig:CFO}
    \end{minipage}
    \hspace{0em}
    \begin{minipage}{0.265\linewidth} 
        \centering
        \includegraphics[width=\linewidth]{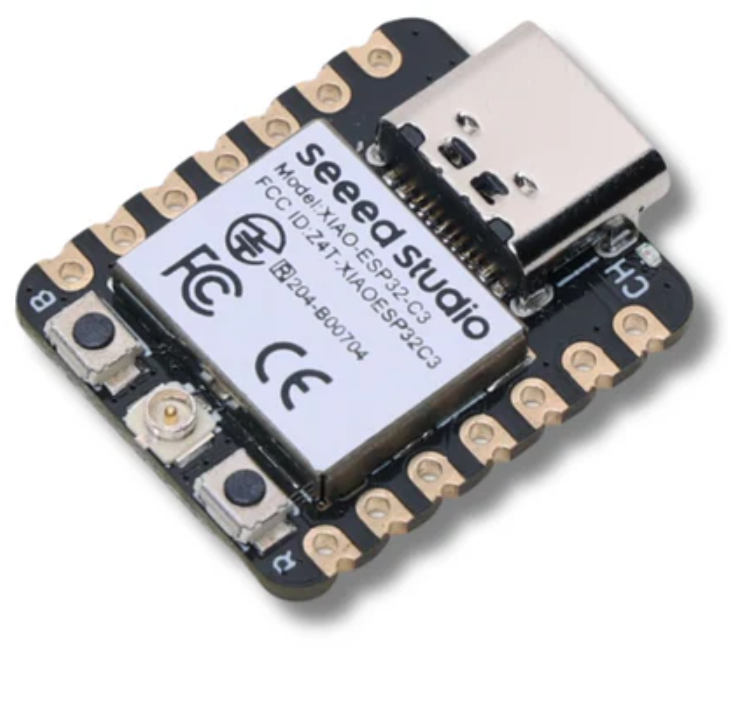}
        \SetCell[c=1]{c} {\small \text{(b) ESP32C3}}
        \label{fig:IQs}
        \vspace{0.1em}
        \includegraphics[width=\linewidth]{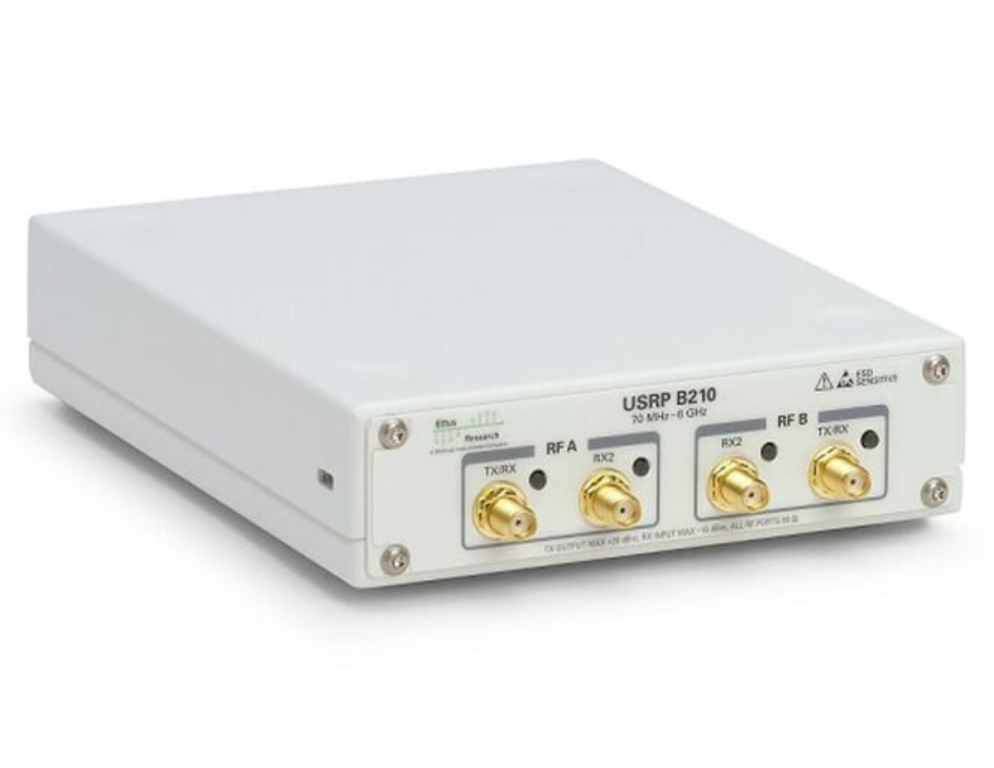}
        \SetCell[c=1]{c} {\small \text{(c) USRP B210}}
        \label{fig:DC}
    \end{minipage}
    \caption{Testbed setup and hardware}
    \label{fig:outdoorsBLE}
\end{figure}

We collected data via both wireless and wired transmissions using channel Ch1. In the wireless collection, raw IQ signals were captured outdoors using four different location scenarios, Loc1, Loc2, Loc3 and Loc4, with the distance between the receiver and the tested devices being 1m, 1.5m, 2m, and 3m, respectively (see Fig.~\ref{fig:outdoorsBLE}(a)).
For all scenarios, we used line of sight (LoS) transmissions and all devices were positioned at the same level. All transmitters' and receiver's antennas were oriented vertically to ensure maximum antenna gain.
The wired data collection took place in the lab, where a coaxial cable was used to connect the devices.

\subsection{Different Channels Under the Same Environment}
\label{sub:DCSE}
In this setup, data were collected using Rx1 in a wired setting, where only the channel index was varied. This was done to isolate the channel effect from the effects of the surrounding environment. When a specific channel is hardcoded into the BLE device, the header field associated with this channel and cyclic redundancy check (CRC) value change accordingly; in other words, this experiment tests not only the impact of frequency hopping but also the effect of changing the frame content.

\subsection{Different Receivers on the Same Channel/Environment}
To evaluate the robustness of our model against receiver change, two different receivers (Rx1 and Rx2) were used to collect data. In this experiment, all signals were transmitted over a wired connection on Ch1.


\section{Performance Evaluation and Analysis}
\label{sec:results}
In this section, we present a comparative analysis of our proposed technique against several baseline methods from the literature. 
To ensure a fair comparison, all methods were evaluated using the same experimental setup, classifier and performance metrics. The classification accuracy, representing the percentage of correctly classified test inputs relative to the total number of test inputs, is used as the metric for our performance evaluation.

\subsection{CNN Architecture}
\label{sub: CNN classifier}
As CNNs have demonstrated effectiveness and robustness in classifying time series signals~\cite{elmaghbub2023needle, kashani2024radio}, the output of the proposed pre-processing technique is passed into a CNN as shown in Fig.~\ref{fig:end to end}. Our CNN architecture has 5 convolution blocks followed by 2 fully connected blocks and an output layer. Each convolution block contains a 1D convolution layer with $F_{i}$ filters and $(1, H_{i})$ kernel size (where $i=1,2,\dots5$ indicates the block index), followed by a batch normalization, leaky Rectified Linear Unit (ReLU) activation function, and max-pooling layer of shape $(1,2)$. Each fully connected block contains a fully connected layer with $N_j$ neurons followed by a leaky ReLU activation function and a drop out layer with rate $P_j$ to reduce overfitting, where $j=1,2$ indicates the fully connected block index. To update our weights, we utilized Stochastic Gradient Descent (SGD) with exponential decaying learning rate. The training was done over $25$ epochs where the batch size was set to $64$. 
Table~\ref{tab:nn_structure} illustrates the exact parameters we used in CNN implementation using the Tensorflow framework. 
\renewcommand{\arraystretch}{1.2}
\begin{table}[ht]
    \centering
    \caption{CNN Architecture Details}
    \begin{tabular}{@{}ll@{}}
        \toprule
        \textbf{Parameter} & \textbf{Value} \\ \midrule
         Number of Filters ($F$) & $\{64,64,96,128,96\}$\\
        Kernel Sizes ($H$) &  $\{48,8,6,5,6\}$ \\
        Number of Neurons ($N$) &  $\{1024, 512\}$ \\
        Dropout Rates ($P$) &  $\{0.5, 0.6\}$ \\
        Initial Learning Rate & $0.05$ \\
        Decay Steps/Rate & $600/0.75$ \\ \bottomrule
    \end{tabular}
    \label{tab:nn_structure}
\end{table}
\subsection{Baseline Fingerprinting Methods}
We compare \TPD{} with two recently proposed methods: 
\begin{itemize}
    \item {\bf Mbed~\cite{jagannath2023embedding}:} Although MBed was originally proposed for RF fingerprinting classification using entire Raw IQ frames, our experiments show that using just the transient and preamble parts of the frame yields higher and more consistent accuracy. Consequently, the  refined approach can be defined as: 
    \begin{equation}
    \mathbf{Mbed}[n] = \begin{bmatrix}
        |\mathbf{x}[n]| \\
        \angle \mathbf{x}[n] \\
        PSD(\mathbf{x}[n])
    \end{bmatrix},
    \quad n = 0,1, \dots L
    \end{equation}
    with $L$, $PSD(\cdot)$ and $|\cdot|$ representing the size of the transient and preamble portions, power spectral density operator and absolute value operator, respectively. 

    \item {\bf TP~\cite{kashani2024radio}:} This method uses the raw Transient and Preamble (TP) of the IQ frames as an input to the classifier, but has not been evaluated when training and testing are done using data obtained from different channels or across different environments.
    
    
\end{itemize}

In addition, we compared the three methods against using raw IQ as the input to the classifier. This will serve as a baseline for all three methods. 

\subsection{Adaptation to Changes in the BLE Channel}
\begin{figure}[t]
    \centering
    \subfigure[Raw IQ]
    {\includegraphics[width=0.99\linewidth]{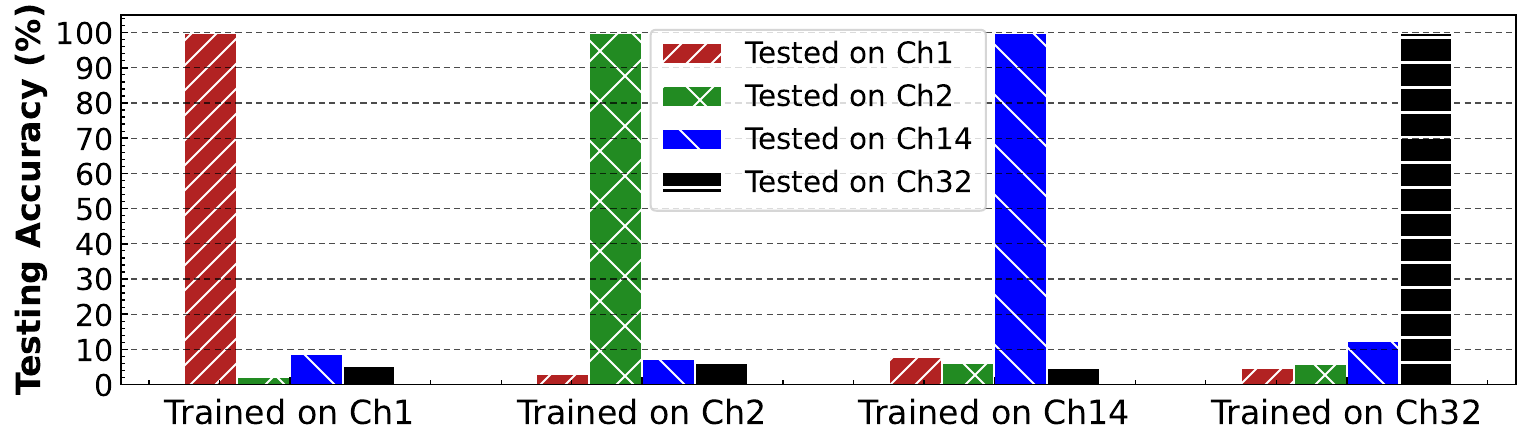} 
    \label{RawIQ}}
    \subfigure[TP]{\includegraphics[width=0.99\linewidth]{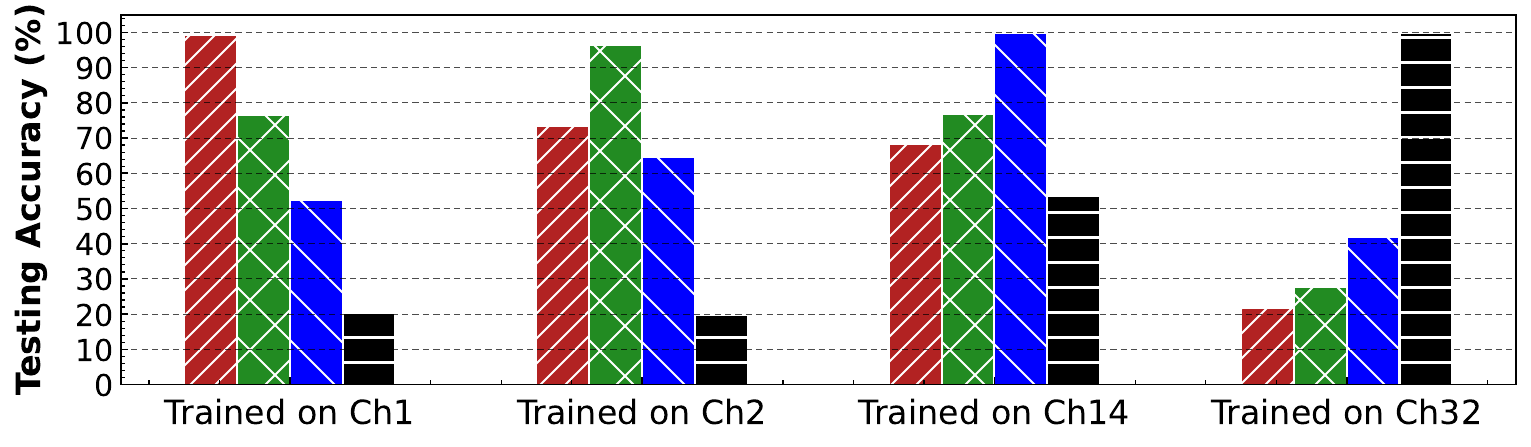} \label{BS2}}
    \subfigure[Mbed]{\includegraphics[width=0.99\linewidth]{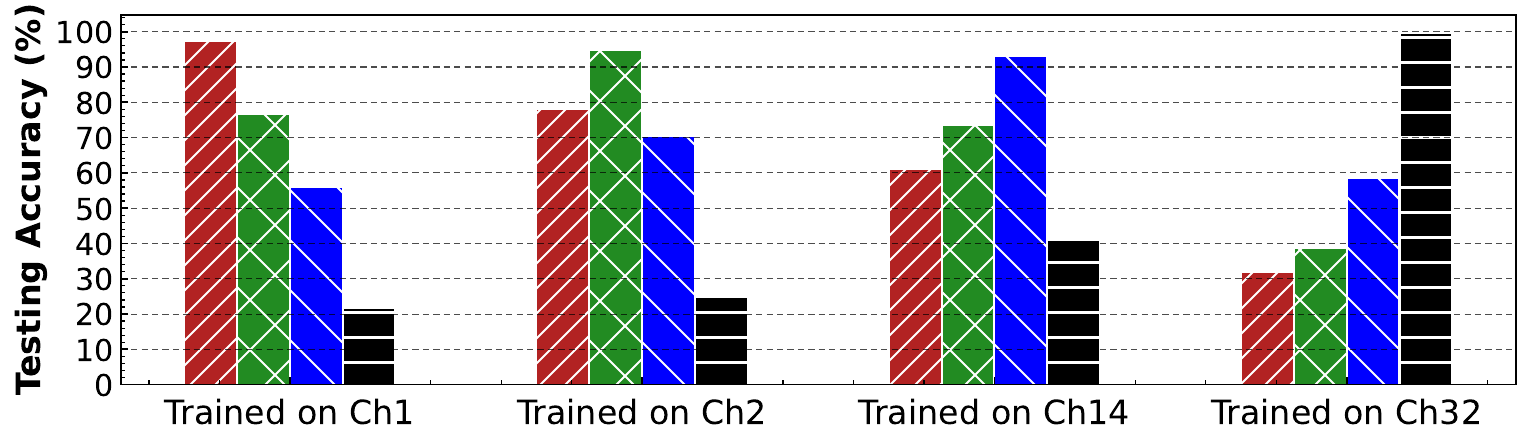} \label{Mbed}}
    \subfigure[\textbf{TPD (ours)}]{\includegraphics[width=0.99\linewidth]{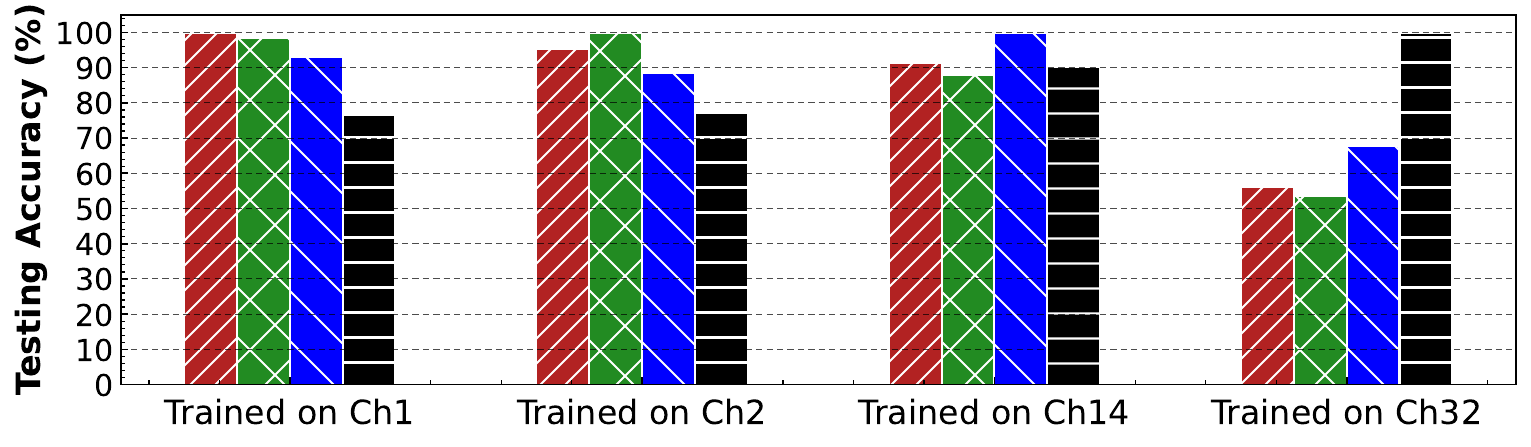} \label{Ours}}
    \caption{Testing accuracy when training and testing across different channels, all in wired environment and using Rx1 for data collection}
    \label{fig:Across Freq}
\end{figure}

Figs.~\ref{fig:Across Freq}{{(a-d)}} compare the classification accuracy of the studied techniques for 31 devices, trained and tested across different channels using data collected from the same receiver, Rx1, and under the same wired environment. Figs.~\ref{RawIQ} and~\ref{BS2} show the results when using the entire raw IQ frame (Fig.~\ref{RawIQ}) and a shorter version containing the transient and the preamble (TP) only (Fig.~\ref{BS2}) as an input to the classifier. 
Even though both techniques are based on the same input, the Raw IQ data based technique drops to less than $10\%$ accuracy when trained and tested on different channels/frame contents, demonstrating that the classifier overfits to the training content (PDU). In contrast, the latter achieves a better accuracy, as it is trained on the transient and preamble portions, which are independent from PDU contents. 

The figures show that our \TPD{} technique (Fig.~\ref{Ours}) outperforms the other proposed methods (Figs.~\ref{BS2} and~\ref{Mbed}), offering 20\% to 58\% higher accuracy. Additionally, although Mbed is relatively expensive to compute, it does not even outperform TP as it relies on the signal phase as one of the primary inputs to the CNN classifier. The phase can be highly affected by the carrier frequency (channel index), causing a significant drop in classification accuracy when the channels are changed.
On the other hand, \TPD{} mitigates channel effects by estimating the rate at which the signal phase changes, effectively minimizing the impact of static phase shifts induced by the communication channel.
Another general trend we observed across all studied techniques is that the classification accuracy decreases as we move away, during testing, from the channel used to train the model.
\subsection{Adaptation to Changes in the Environment}
\begin{figure}
    \centering
    \includegraphics[width=\linewidth]{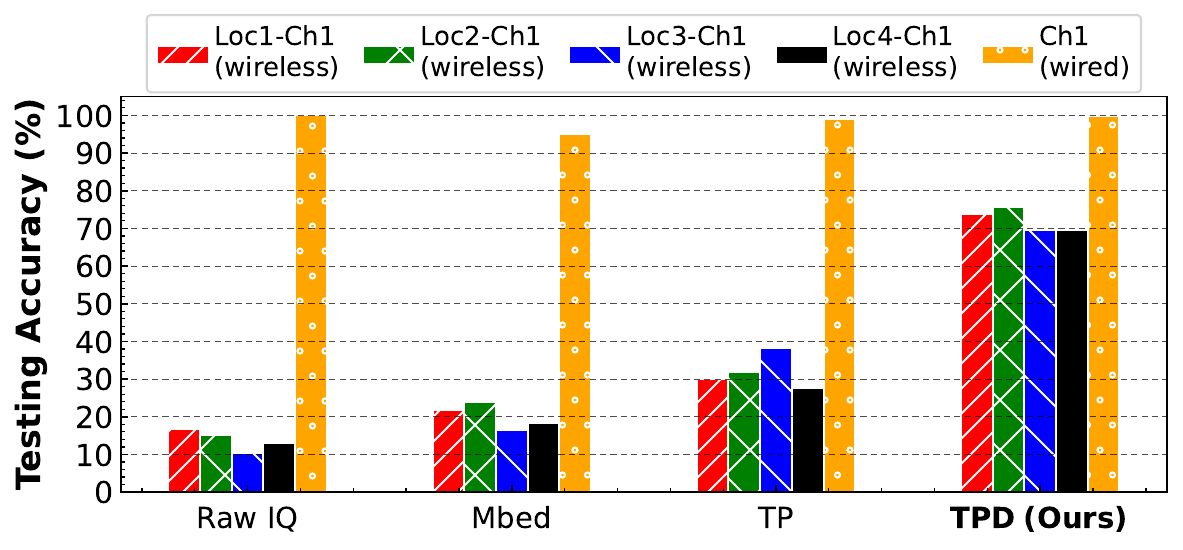}
    \caption{Testing accuracy when training on wired data, but testing on wireless data with varying locations: All data was collected using receiver Rx1 and channel Ch1.}
    \label{fig:Across envs}
\end{figure}

Fig.~\ref{fig:Across envs} shows the classification accuracy when training is done on wired data but testing is done on wireless data under varying devices' locations. All training and testing data was collected using receiver Rx1 and channel Ch1.
The figure reveals that our method achieves a stable, high classification accuracy over 31 devices, reaching roughly between 70\% and 75\% when training is done on wired data but testing is done on wireless data, that remains consistent even when changing the locations of the devices. 
The results show that our method increases the accuracy by around 40\% to 45\% compared to TP, and by around 50\% compared to Mbed. 

\begin{figure}[t]
    \centering
    \includegraphics[width=\linewidth]{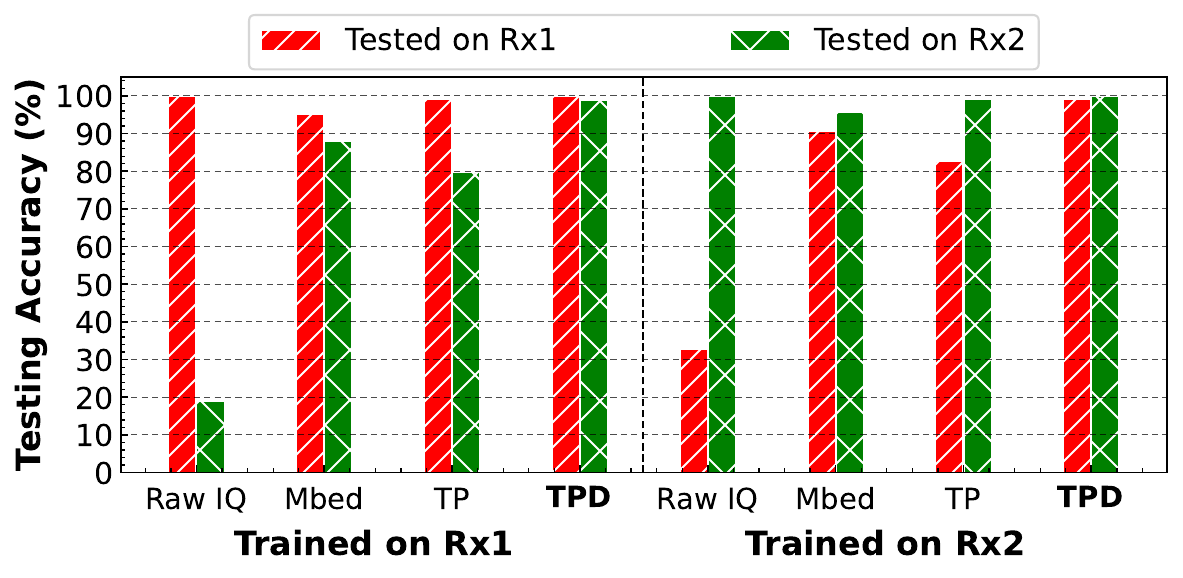}
    \caption{Testing accuracy when training and testing data are collected over a wired setup on Ch1 using different receivers.}
    \label{fig:Across Rxs}
\end{figure}

\begin{figure}
    \centering
    \includegraphics[width=\linewidth]{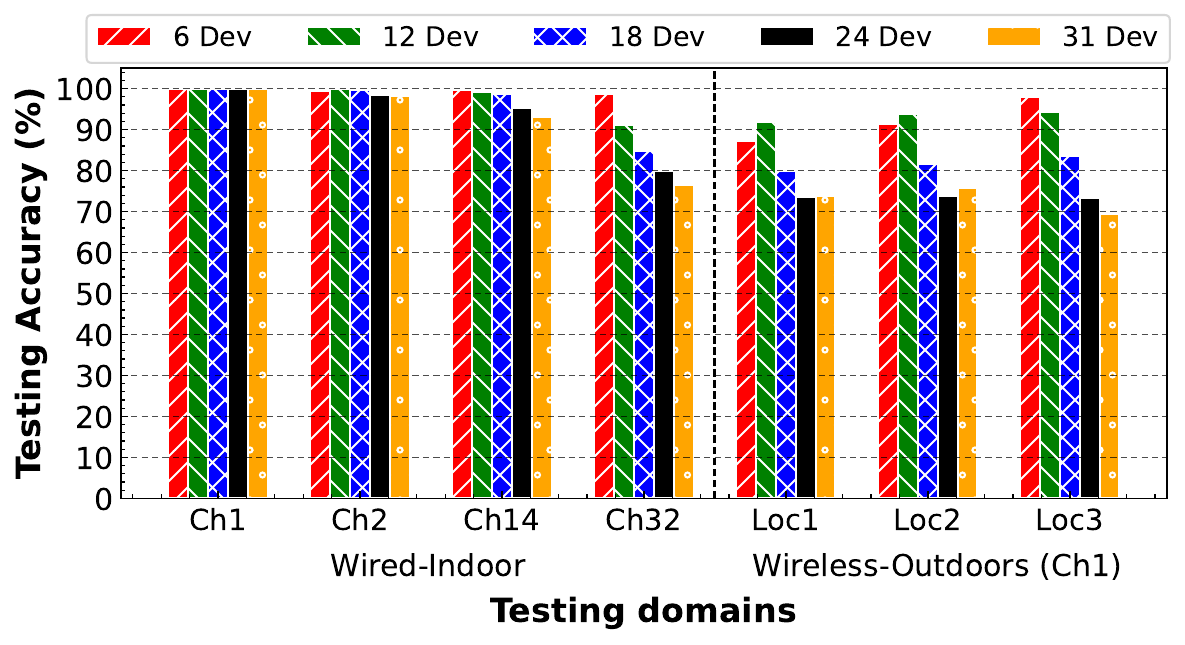}
    \caption{Classification accuracy when varying the number of tested devices: all trained on data collected using wired connection and on Ch1.}
    \label{fig:scalability}
\end{figure}

\begin{figure*}
    \centering
    \subfigure[Tested on Ch2 (wired)]{\includegraphics[width=0.324\linewidth]{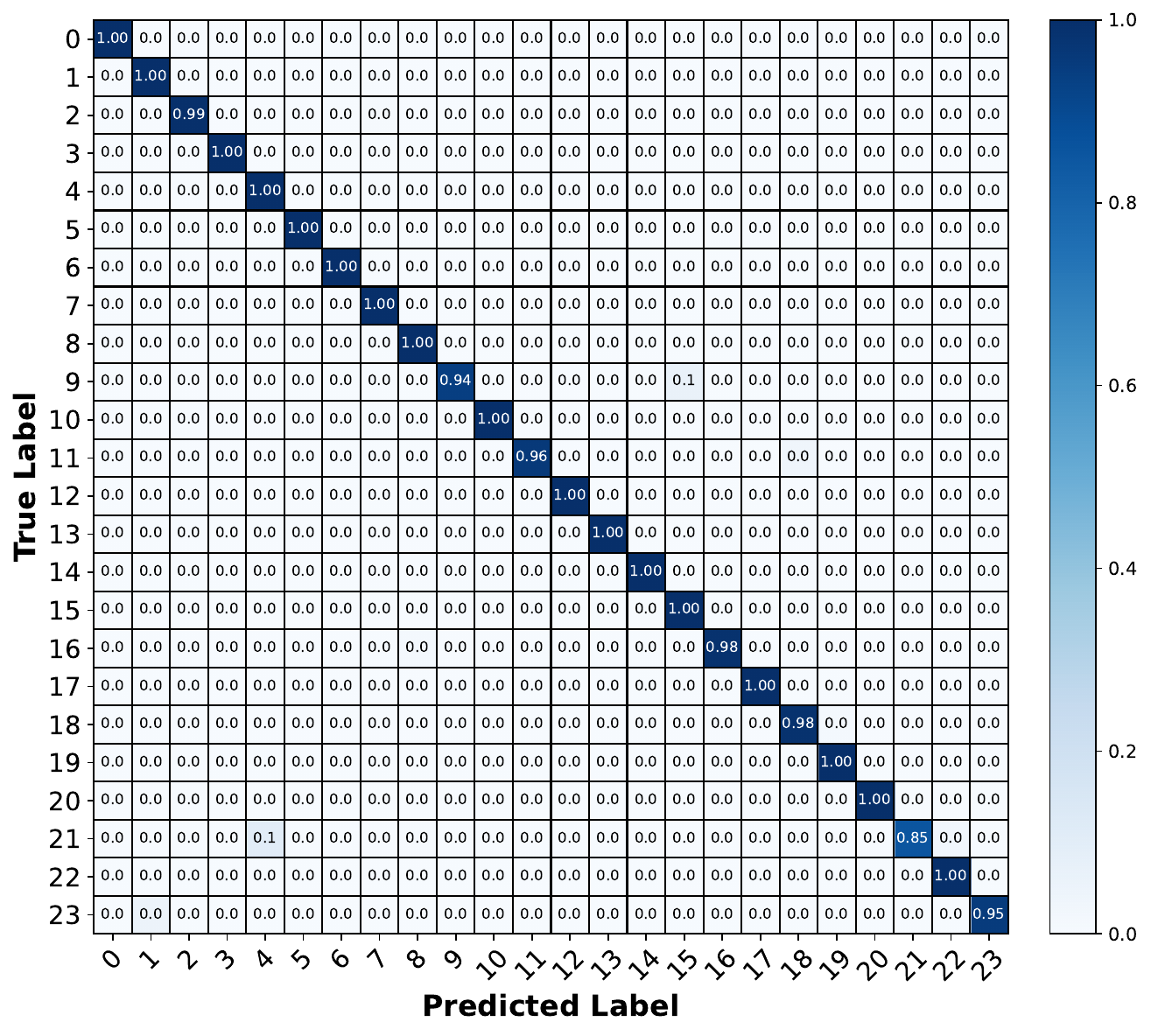}}
    \subfigure[Tested on Ch14 (wired)]{
    \includegraphics[width=0.324\linewidth]{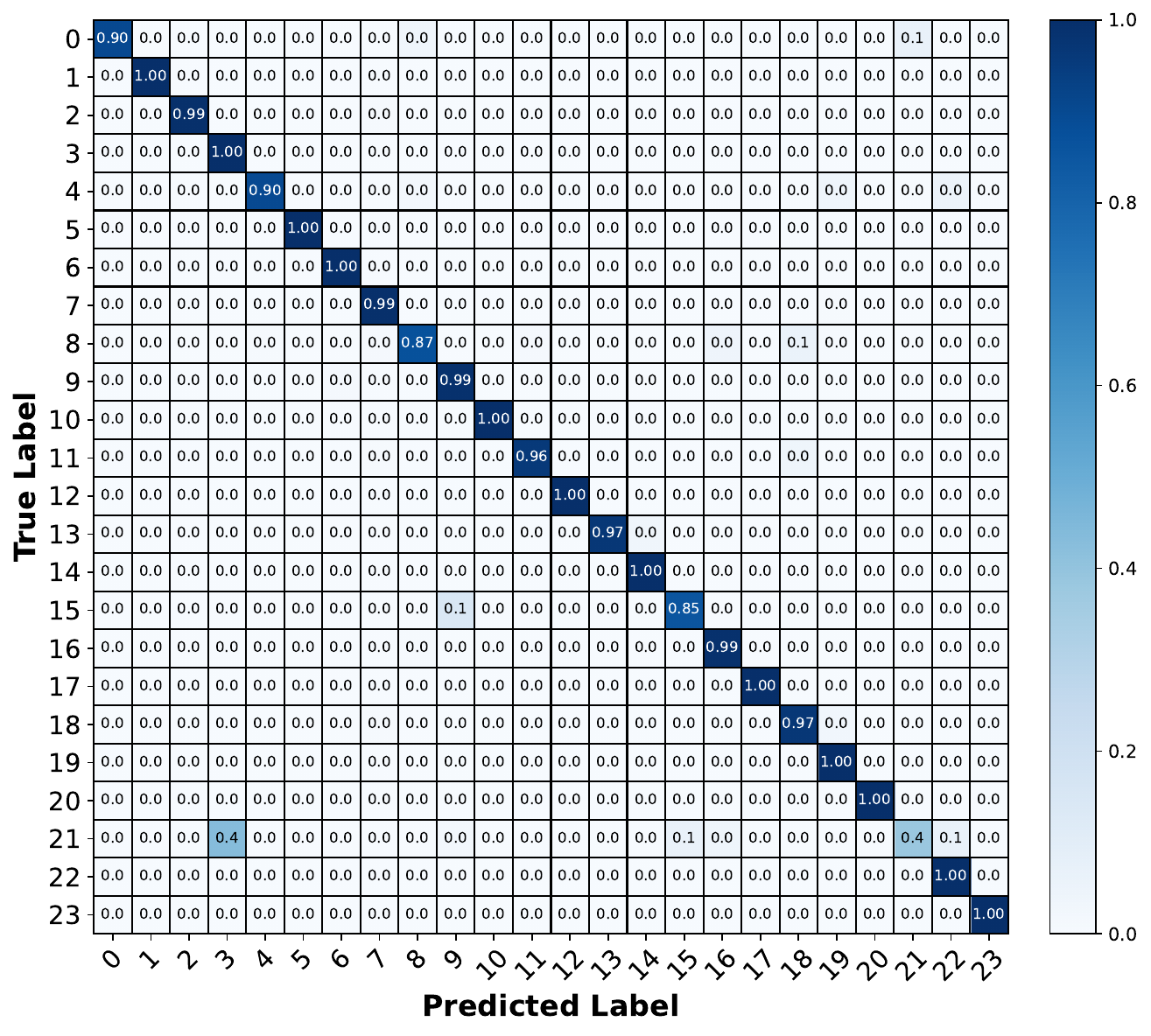}}
    \subfigure[Tested on Ch32 (wired)]{
    \includegraphics[width=0.324\linewidth]{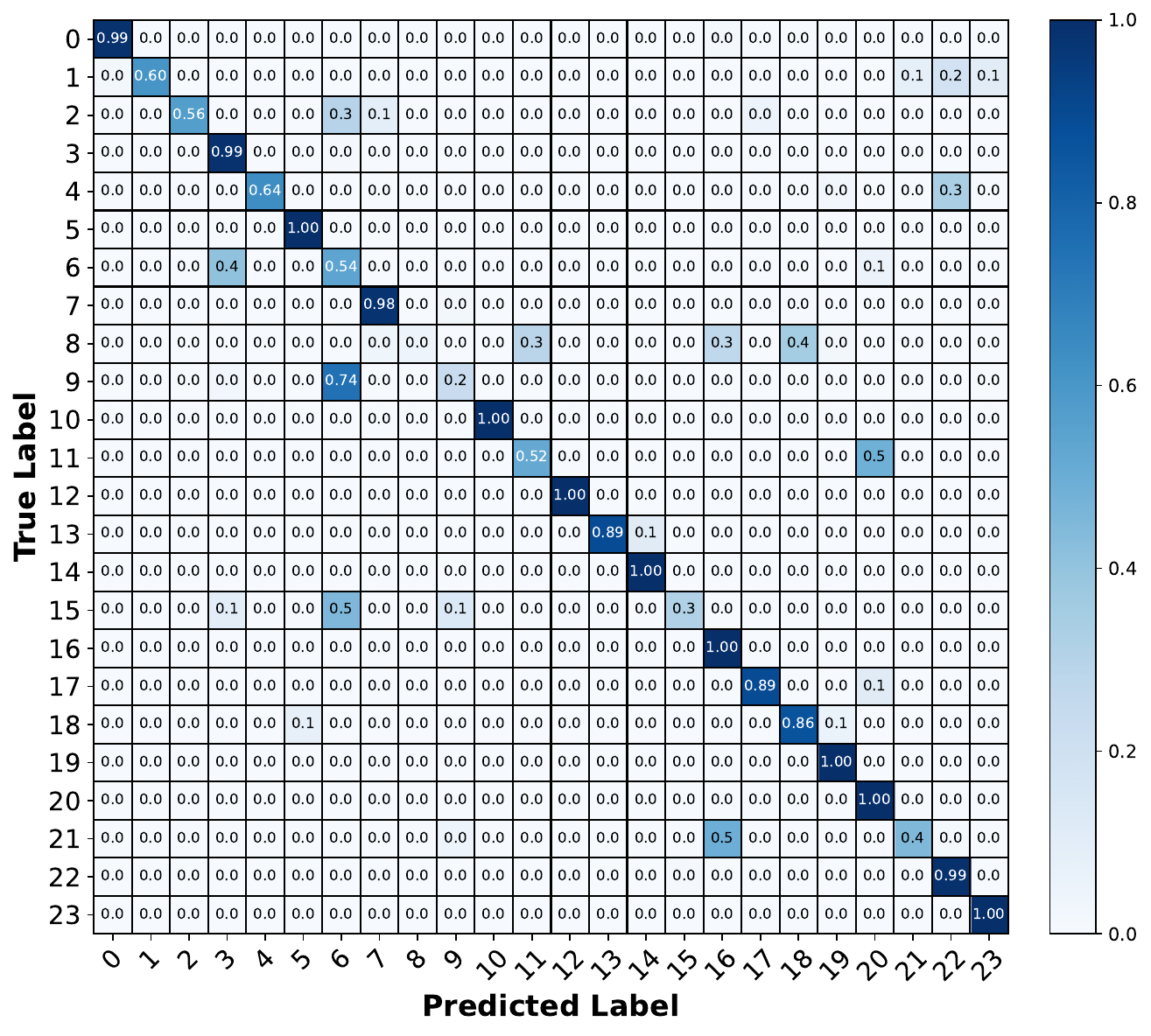}}
    \subfigure[Tested on Loc1-Ch1 (wireless)]{\includegraphics[width=0.324\linewidth]{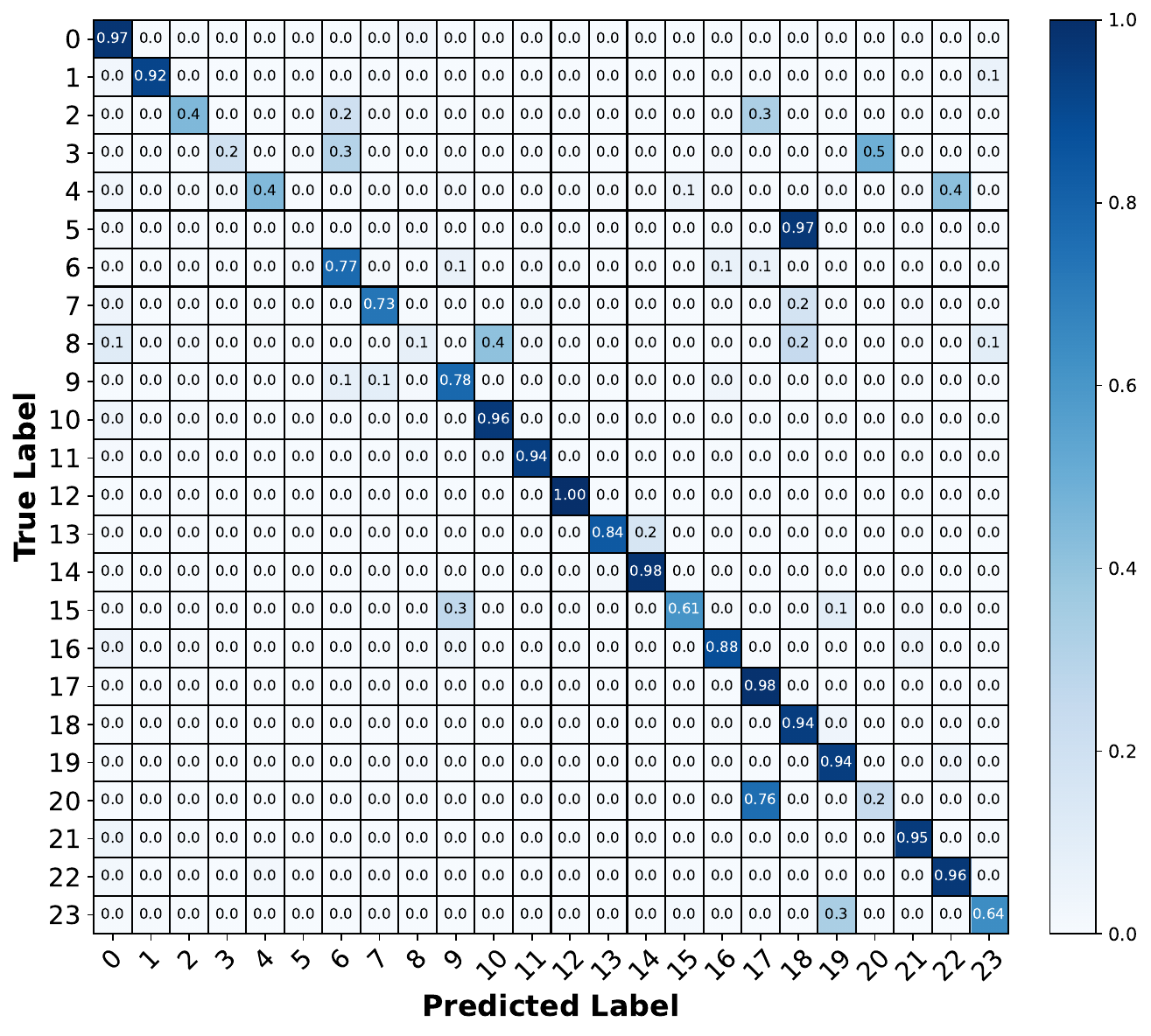}}
    \subfigure[Tested on Loc2-Ch1 (wireless)]{\includegraphics[width=0.324\linewidth]{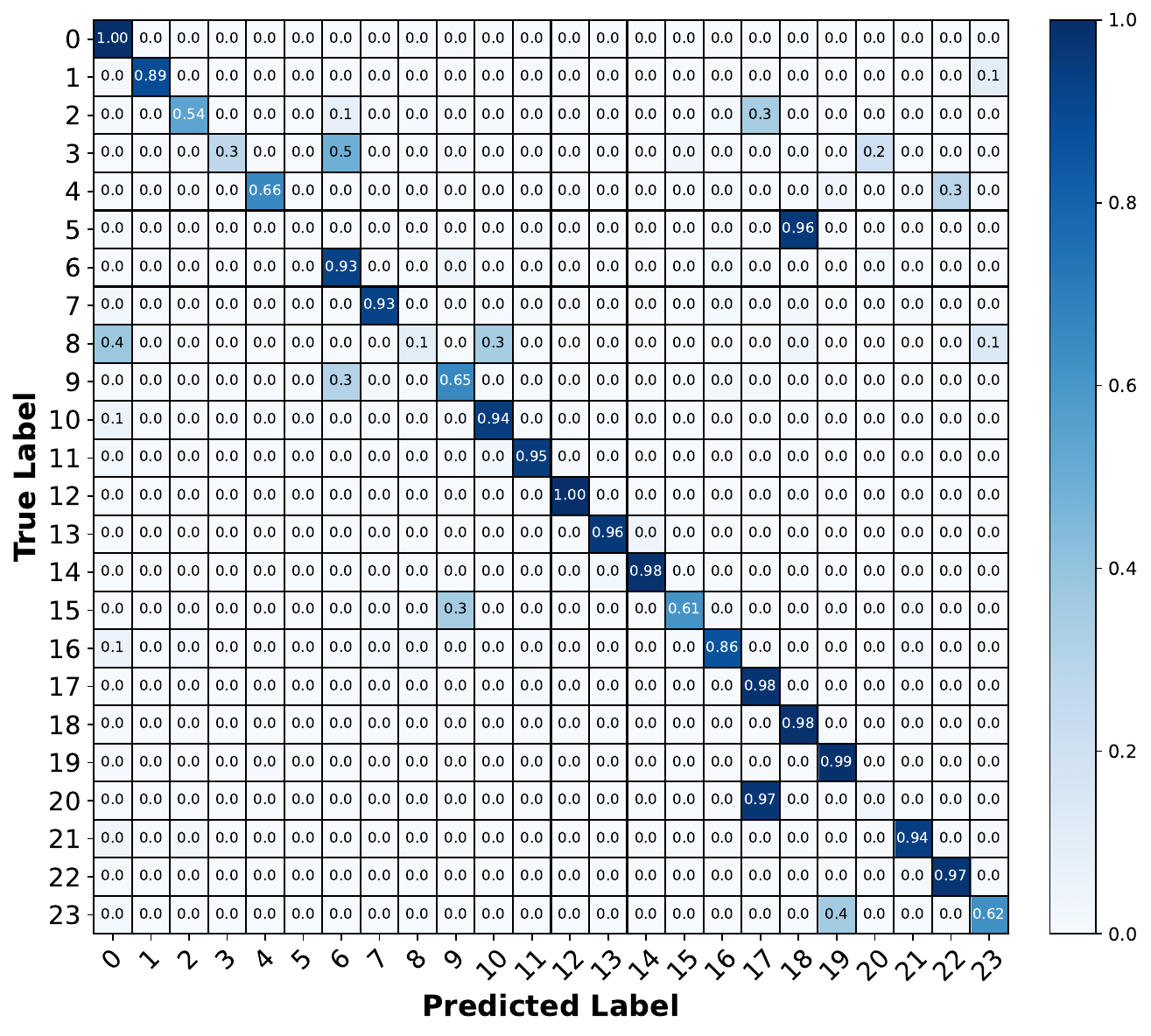}}
    \subfigure[Tested on Loc3-Ch1 (wireless)]{\includegraphics[width=0.324\linewidth]{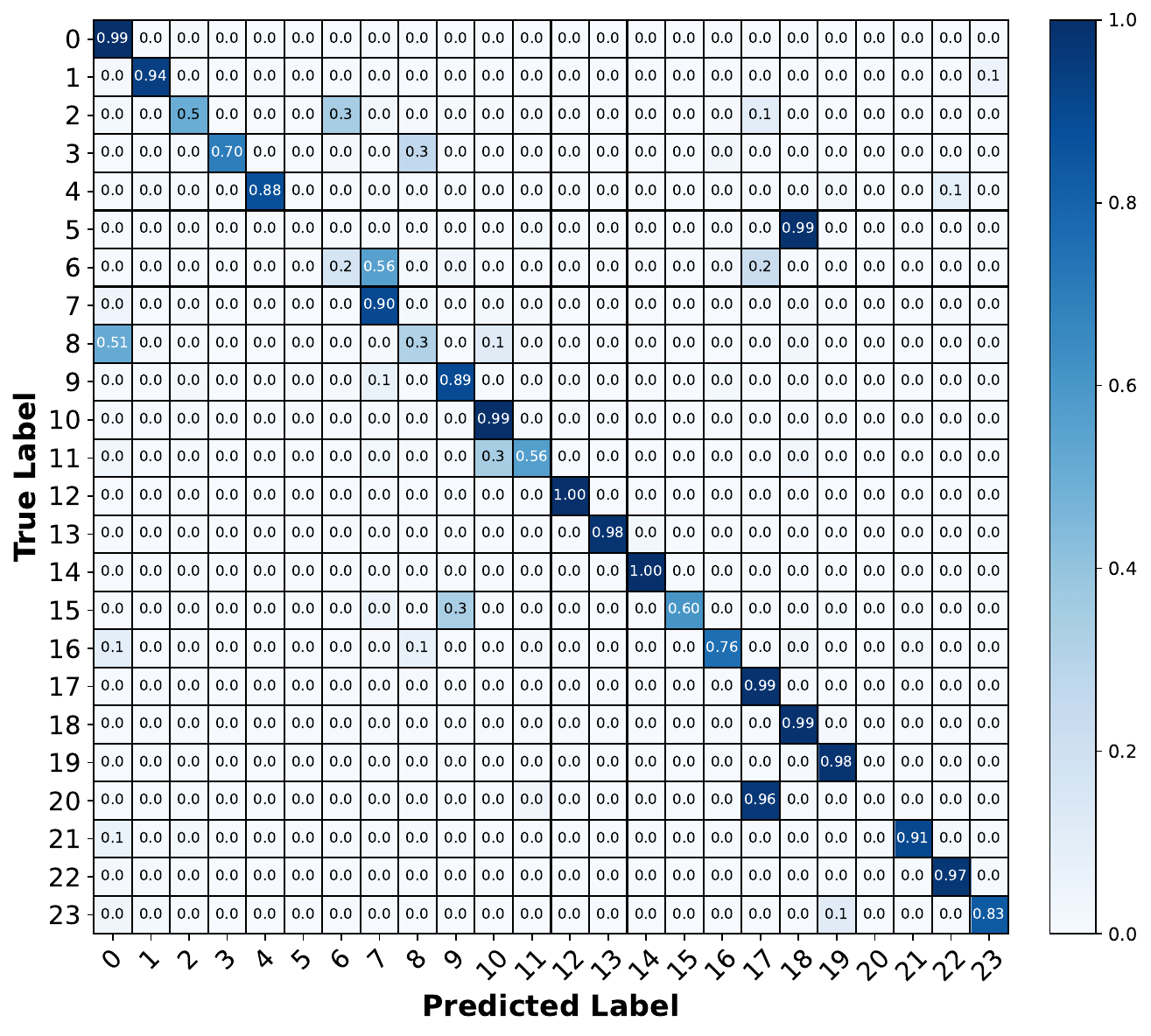}}
    \caption{Confusion matrices when training is done on data collected on wired connection using Ch1.}
    \label{fig:Confusion}
\end{figure*}

\subsection{Adaptation to Changes in the Receiver}
Fig.~\ref{fig:Across Rxs} shows the results when training data is collected on one receiver but the testing data is collected on another receiver.  
It indicates that \TPD{} adapts better than the other two techniques, maintaining consistent accuracy of about 99\% regardless of whether the same or a different receiver is used. On the other hand, changing receivers leads to a drop in accuracy from about 95\% to 88\%  and from about 98\% to 81\% for Mbed and TP, respectively.

\subsection{Execution Time Analysis}

Beyond classification accuracy, it is equally important to evaluate the computational efficiency of different feature representations. To this end, we report the \emph{empirical execution time} of each method, measured in terms of preprocessing, training, and inference.
Table~\ref{tab:comparison} summarizes the results, where training time was measured over 10 epochs, preprocessing time captures the cost of transforming raw I/Q samples into the corresponding feature domain, and inference time quantifies the latency of predicting class labels for unseen signals. For Raw IQ and TP, preprocessing is defined simply as stacking the real and imaginary parts of the signal into the channel dimension before feeding them into the classifier. Bold values highlight the best performance in each category.

The results indicate that Raw IQ has the highest training time $(111.49 s)$, which is nearly four times slower than the proposed representations. Among the feature-based methods, TP achieves the lowest preprocessing overhead $(0.000097 s)$, while \TPD{} offers the fastest overall training $(29.64 s)$ and inference $(0.56 s)$. Mbed exhibits slightly higher preprocessing cost due to FFT computation, but remains competitive in training and inference compared to TP. Overall, TP and \TPD{} demonstrate significant efficiency gains over Raw IQ, highlighting the benefit of lightweight feature transformations in reducing both training and inference latency for practical deployment.

\begin{table}
    \centering
    \caption{Execution Time Comparison of Baselines and Proposed Technique (in seconds)}
    \label{tab:comparison}
    \begin{tabular}{l|ccc}
        \toprule
        \midrule
        \textbf{Method} & \textbf{Processing time} & \textbf{Training time} & \textbf{Inference time} \\
        \midrule
        {Raw IQ} & 0.000139 & 111.486 & 1.40 \\
        \midrule
        {TP} & \textbf{0.000097} & 30.442 & 0.58 \\
        \midrule
        {Mbed} & 0.000521 & 33.8 & 0.61 \\
        \midrule
        {TPD} & 0.000963 & \textbf{29.643} & \textbf{0.56} \\
        \midrule
        \bottomrule
    \end{tabular}

\end{table}

\subsection{Scalability of \TPD{} with the Number of Devices}
We also show in Fig.~\ref{fig:scalability} the impact of the number of devices using subsets of 6, 12, 18, 24 and 31 devices. For each subset, all models were trained on Ch1/Rx1 (wired) setup. 
The figure shows a monotonic decrease in accuracy as the number of devices increases. Notably, the farther away the testing channel is from the training channel, the greater the decrease becomes for larger subsets.\\
To further investigate the impact of domain shift, we present the confusion matrices in Fig.~\ref{fig:Confusion} for 24 devices. As shown in subfigures~\ref{fig:Confusion}(a-c), the misclassification rate increases as we move further away from the training channel (Ch1). Additionally, the wireless environment, illustrated in subfigures~\ref{fig:Confusion}(d-f), introduces a biased shift in the impairments of certain devices. This shift makes the receiver misclassify device 20 as 17 and device 5 as 18 with high confidence. We believe this issue arises from the fading nature of wireless channels, which distorts the signal characteristics and leads to misclassifications. To mitigate these effects, more advanced techniques, such as equalization, may be necessary to compensate for channel-induced distortions and improve classification accuracy.




\section{Conclusion}
\label{sec:conclusion}

We propose \TPD{}, a feature extraction technique that effectively extracts device signatures from RF signals transmitted by BLE devices. 
Through experimental datasets, we demonstrate that the proposed \TPD{} representation of IQ signals significantly enhances the scalability, robustness, and generalizability of deep learning models compared to using raw IQ data and two baseline methods, underscoring its effectiveness and advantages.

\section{Acknowledgment} 
This work is supported in part by NSF Award 2350214.

\bibliographystyle{IEEEtran}
\bibliography{References}

\end{document}